\documentclass{revtex4}
\usepackage{epsfig}

\newcommand{\be}{\begin{equation}}
\newcommand{\ee}{\end{equation}}
\newcommand{\bea}{\begin{eqnarray}}
\newcommand{\eea}{\end{eqnarray}}

\begin{document}

%\begin{frontmatter}

\title{On some common misconceptions regarding the "Ergodic Hierarchy".}
\author{Henk van Beijeren} 
\address{Institute for Theoretical Physics\\
Utrecht University\\ 
Leuvenlaan 4, 3584 CE Utrecht, The Netherlands}

\date{\today}

\begin{abstract}
The well-known ergodic hierarchy of sheerly ergodic, mixing, Kolmogorov and 
Bernoulli systems, with each next level supposedly encompassing the previous 
one, is shown to be too simplistic in its usual formulation. 
A K-system can be sheerly ergodic and sometimes may be reduced to a sheerly 
mixing system by some simple projection. More precise characterizations of 
ergodic properties of dynamical systems should start out from a consideration of 
the full Lyapunov spectrum.
\end{abstract}
%\begin{keyword} 
\pacs {05.45.-a, 05.20Jj }
%\end{keyword}
%\end{frontmatter}

\maketitle

\section{Introduction}

In ergodic theory it has been common to define a hierarchy of degrees of 
ergodicity \cite{arnold}. The weakest property is sheer {\it ergodicity}. This 
implies that, 
under some given dynamics, the time average of any Lebesgue integrable function 
on phase space equals an average over some properly chosen subset ${\mathcal S}$ 
of 
phase space. If the dynamics leaves volumes in phase space unchanged, as is the 
case for Hamiltonian dynamics, this phase space average simply assigns weights 
proportional to phase space volume to various parts of this subset. If the 
dynamics is phase space contracting one has to choose an {\it attractor} as 
subset and use the SRB-measure for its weighting, i.e.\ each part of the subset 
has to be weighted by its asymptotic phase space contraction factor. The next
property in the ergodic hierarchy is {\it mixing}. This implies that under the 
dynamics the image of any subset ${\mathcal A}$ of non-zero measure becomes 
uniformly distributed on  ${\mathcal S}$, i.e.\ the measure of its intersection
with an arbitrary other subset  ${\mathcal B}$ of non-zero measure approaches 
$\mu({\mathcal 
A})\mu({\mathcal B})/\mu({\mathcal S})$ as time increases. Mixing implies 
ergodicity, but not the other 
way around; therefore mixing is a stronger property than sheer ergodicity. In 
fact one may distinguish between {\it strong mixing}, where the uniformity 
property
holds pointwise in time, and {\it weak mixing} where it only holds on average 
over long stretches of time. The next property in the ergodic hierarchy is that 
of being a {\it Kolmogorov system} or K-system. The original definition 
of this is rather involved, but for my present purposes it suffices that 
K-systems
are systems with positive Kolmogorov-Sinai entropy (or KS-entropy). For systems
without escape, i.e.\ with dynamics mapping all points of ${\mathcal S}$ onto 
${\mathcal 
S}$, according to Pesin's theorem this is equivalent to the requirement that the 
system has at least one positive Lyapunov exponent, or in other words, is 
chaotic.
It is commonly assumed that being a K-system implies mixing and therefore is a 
stronger property. An even stronger property, the strongest in the ergodic 
hierarchy, is for a system being a Bernoulli system. This is most easily defined
for maps, rather than flows (for the latter one therefore may take recourse to 
time slices or to Poincar\'e maps). It implies that ${\mathcal S}$ may be 
decomposed 
into a discrete collection of subsets ${\mathcal A}_i$ with the following 
property: 
for any subset of all points on ${\mathcal S}$ with a given history, i.e.\ with 
a 
given sequence $i_{-1},i_{-2},\cdots$ of subsets 
% ${\mathcal A}_{i}$ 
visited at previous mappings, the fraction of points ending up in ${\mathcal 
A}_{i}$ 
at the next mapping is simply proportional to the measure of this subset. In 
other words, without additional information on where the phase point is or came 
from, the outcome of the mapping is like determined by pure chance, irrespective 
of what happened before.

Here I will argue that in fact this hierarchy should not be taken too 
litterally. A K-system may be non-mixing and even if it is mixing, its ergodic 
behavior
in certain respects need not be stronger than that of a merely mixing system.

\section {Mixed behaviors}

%\label{sec:mix}
A standard example of a map exhibiting the Bernoulli property is the baker map,
defined on the unit square as
\be
B(x,y)=(2x\  mod\  1,y/2+ int(2x)),
\ee
with $int(x)$ the integer part of $x$.This map is area conserving and has 
Lyapunov exponents $\pm\log 2$, with corresponding eigenvectors in tangent space 
$(1,0)$ and $(0,1)$. It satisfies a Bernoulli scheme with respect to the 
subdivision of the unit square into the two subsets $0\le x<1/2$ and $1/2\le 
x<1$. It may be extended to a map on the unit cube of the form
\be
M(x,y,z)=(2x\  mod\  1,y/2+ int(2x),f(z)),
\ee
with $f(z)$ mapping the unit interval onto itself. This map is still Bernoulli, 
as it still satisfies a Bernoulli scheme with respect to the subdivision, now of 
the unit cube, into the subsets  $0\le x<1/2$ and $1/2\le x<1$. It still has a 
Lyapunov exponent $\log 2$, so it obviously is still a K-system, but whether it 
is mixing depends on the nature of the function $f$. If $f$ is merely ergodic,
e.g.\ a shift $mod\ 1$ over an irrational number $\alpha$, the map $M$ is not 
mixing. So the statement that the property of being a K-system implies mixing, 
obviously is not true. In this case one may say that $M$ combines the property 
of being Bernoulli when projected on the $x-y$ plane, with that of being merely 
ergodic when projected on the $z$-axis. Therefore, to me it also seems dubious
to state in general that K-systems are more strongly ergodic than mixing 
systems. In the present example I would say that the system is no 
more ergodic than the one defined by the one-dimensional mapping $f$ alone.

One may object now that the present example is very atypical because it can be 
factorized into a two-dimensional mapping that is Bernoulli and a 
one-dimensional mapping that is merely ergodic. It is not hard though to extend 
this example into non-factorizable mappings, e.g.\
\bea
M_1(x,y,z)=&&(2x+z\  mod\ 1,y/2+int(2x),z+\alpha \nonumber \\
&&+(2x+z\  mod\ 
1)+y/2+int(2x)-x-y),
\eea
which acts like a shift map to which an $x$ and $y$ dependent constant is added.
Going to higher dimensions one may, for example in d=4, devise mappings that 
asymptotically reduce to a baker map in the $x-y$ plane times a quasi-periodic 
map on some curve in
the remaining coordinates. In this case the decoupling of the two maps may 
easily be designed to hold only asymptotically on the attractor. No doubt
increasingly more complicated scenario's may be designed with increasing 
dimensionality.

\section{Discussion}

For the general case of a mapping or flow in a high-dimensional phase space the
ergodic properties of the dynamics on a given ergodic component will first of 
all depend on the spectrum of Lyapunov exponents. For simplicity, let us 
restrict 
ourself to symplectic maps or flows, for which the spectrum is odd (a Lyapunov 
exponent 
$\lambda$ always comes together with $-\lambda$). Then, if there are no zero 
exponents, in the case of maps, or, in the case of flows, just a single pair 
related to displacements in the direction of the flow, one may say the system 
fully has the character of a K-system. No projections are possible in directions 
where the dynamics is merely mixing or even merely ergodic. Whether the system 
is also Bernoulli is a separate problem that has to be decided case by case. If 
there are more zero Lyapunov exponents, one should decide first of all what is
the character of the behavior of the dynamics in the corresponding directions.
This may be hard to decide, especially if the dynamics in these directions 
cannot be decoupled from that in the remaining ones.

Without further specifications the usual ergodic hierarchy remains a qualitative
characterization of increasingly randomizing and chaotic types of dynamics.
As such it is extremely useful. But for fixing a rigorous sequence in which each 
next stage implies the previous one, obviously more precision is needed.

\acknowledgements
HvB acknowledges support by the Mathematical physics program of
FOM and NWO/GBE.

\bibliographystyle{prsty}

\end{document}